\documentclass[preprint,amsmath,amssymb,11pt,english,aps,showpacs,showkeys,pra]{revtex4-2}
\usepackage{graphicx}
\usepackage{graphics}
\usepackage{amsmath}
\usepackage{dcolumn}
\usepackage{amssymb}
\usepackage{bm}
\usepackage[utf8]{inputenc}

\begin{document}
\title{Quantum Holonomy based in a Kaluza-Klein description for defects in $C_{60}$ fullerenes}
\author{Everton Cavalcante}
\email{Electronic address: everton@ccea.uepb.edu.br}
\affiliation{Centro de Ci\^encias Exatas e Sociais Aplicadas, Universidade Estadual da Para\'{\i}ba, Patos, PB, Brazil}
\author{Claudio Furtado}
\email{Electronic address: furtado@fisica.ufpb.br}
\affiliation{Departamento de F\'{\i}sica, Universidade Federal da Para\'{\i}ba, Caixa Postal 5008, 58051-970, Jo\~ao Pessoa, PB, Brazil}


\begin{abstract}

In this paper, we discuss a new way to get a quantum holonomy around topological defects in $C_{60}$ fullerenes. For this, we use a Kaluza-Klein extra dimension approach. Furthermore, we discuss how an extra dimension could promote
the formation of new freedom degrees which would open a discussion about a possible qubits computation.

\end{abstract}


\pacs{F,G,H}

\maketitle

\section{Introduction}\label{secI}
Nowadays, great efforts in theoretical physics are devoted to study models with extra dimensions. Extra dimensions concept have a long story that began by the decade of 1920 when  Kaluza \cite{Kaluza} and Klein \cite{Klein} proposed the first of these kind of models.
Within the Kaluza-Klein (KK) approach, the metric tensor is represented by a function of an additional vector and scalar fields, $g_{\mu \nu}= g_{\mu \nu}\left(A_{\mu}(x),\phi (x)\right)$. This change in the metric not only reproduces the field equations for $g_{\mu \nu}(x)$, but also recovers correctly the Maxwell theory for $A_{\mu}(x)$ and the massless Klein-Gordon equation for $\phi(x)$. 
Thus, a new relationship between electromagnetic systems and general relativity in five dimensions has been introduced, and a new branch of research has been founded (for a review on Kaluza-Klein theory see f.e. \cite{Salam, Duff, Love}).
All of these efforts, as we already know, have not yet been able to unite gravity theories and quantum phenomena. However, such efforts have been elucidating possibilities of research in a  many branches of the physics. One of these possibilities has proposed in this work.  

The quantum dynamics for fermions in the presence of topological defects in solids is described by local reference frames, and the spinors transform under infinitesimal Lorentz transformations \cite{Kleinert}. In this way, to preserve a local Lorentz group, we  introduce a covariant derivative $\nabla_{\mu}=\partial_{\mu}+\Gamma_{\mu}(x)$. This one allow to describe a geometric phase on spinor as a function of $\Gamma_{\mu}(x)$ \cite{Mostafazadeh}.
On the other hand, a way to implement a holonomic quantum computation (HQC) was proposed by Zanardi and Rasetti \cite{Zanardi}. Such an idea is based on non-Abelian geometric phases called holonomies \cite{Wilczek, Mostafazadeh}. 
One of the interesting characteristics of HQC is its stability \cite{Kuvshinov}, which inspired a lot of works dedicated to their applications in different quantum systems, such as ion traps \cite{Monroe Meekhof, Pachos 2002}, Josephson-junction devices \cite{Cholascinski 2004}, neutral atoms \cite{Recati 2002}, optical computer \cite{Pachos Chountasis}, conical defects in graphene layers \cite{Bakke Furtado 2009}, nuclear magnetic resonance \cite{Jones Vedral 2000} and nuclear Zeeman-perturbed systems \cite{Teles 2015}.
Besides this, the quantum gates have been realized in terms of operations possessing a purely geometrical nature, with the stability of quantum gates against local perturbations must be emphasized especially \cite{Pachos Zanardi 2001}.
Clearly, in HQC, the states of a quantum system are used for encoding information by means of a holonomy (unitary evolution) for a final state where the solution of a given computational problem can be decoded. This state can be presented as
\begin{equation}
|\psi \rangle= U(\alpha)| \psi_{0} \rangle =e^{-i\int_{0}^{t}E(t')dt'}\Gamma_{A}(\alpha)| \psi_{0}\rangle
\end{equation}
The exponential of the integral here is the dynamical phase, which can be omitted after redefining the zero energy level as $E(0)=0$, the $\Gamma_{A}(\alpha)=\exp \int_{c}A(\alpha)d\alpha $  is the holonomy. The quantity $A(\alpha)$ corresponds to the Berry potential with components $A^{\mu \nu}=\langle \psi^{\mu}(\alpha)| \frac{\partial}{\partial \alpha} | \psi^{\nu}(\alpha) \rangle$ used to describe arising of geometric phases in any non-adiabatic cyclic evolution of a quantum system \cite{Aharonov Anandan}. The most known examples of non-adiabatic cyclic evolutions are the Aharonov-Bohm effect \cite{Aharonov-Bohm effect}, the scalar Aharonov-Bohm effect \cite{scalar Aharonov-Bohm effect}, the Aharonov-Casher effect \cite{Aharonov-Casher effect}, the He-McKellar-Wilkens effect \cite{He-McKellar, Wilkens effect}, and the dual Aharonov-Bohm effect \cite{dual Aharonov-Bohm effect}. 

Recently, two interesting ideas about graphene layers were proposed. 
One of these ideas is the possible implementing a quantum holonomy associated with a defect structure in a conical graphene \cite{Holonomic quantum computation associated with a defect structure of conical graphene, One-qubit quantum gates associated with topological defects in solids}, where the geometric description for a defects in solids proposed for Katanaev and Volovich was used \cite{Katanaev Volovich}. It was demonstrated there how to construct quantum gates depending only of the number of sectors ($n_{\Omega}$) in the graphene layer removed in order to build a particular conical configuration described by a two-dimensional metric. 
However, in \cite{A Kaluza Klein description of geometric phases in graphene}, it has been shown that it is possible to extend  the usual two-dimensional metric describing a topological defect in a graphene layer to a three-dimensional metric by adding an Kaluza-Klein extra dimension to introduce the Fermi-point degree of freedom
within the geometrical phase acquired by wave function of the fermions which go around the apex of a defect in a graphene layer.

In other work developed by us \cite{Everton}, a geometrical model of a fullerene molecule with a $Ih$ symmetry was described. For this, we have combined the well known non-Abelian monopole approach \cite{Maria} and the geometric theory of defects \cite{Katanaev Volovich}. Thus, we have applied a continuum formulation to describe this spherical geometry with a bi-dimensional spherical metric. 
Our aim in this work is to discuss a way to get a holonomy matrix in the $C_{60}$ molecule through a Dirac phase factor method within an extra dimension approach.

This paper is organized as follows. In section \ref{secII}, we give a brief review of the geometrical approach for a $Ih$ symmetrical fullerene molecule. In section \ref{secIII}, we introduce the extra-dimensional approach for topological defects. After that, in section \ref{secIV}, we study the geometrical phases for massive and massless modes of fermions and discuss a new way of obtaining quantum holonomies for the system. Finally, in section \ref{secV}, we present our conclusions.
\section{Geometric approach for topological defects in $C_{60}$ fullerenes}\label{secII}

In 1985 the molecule of fullerene was discovered by Kroto,
Curl, and Smalley \cite{Kroto}. In this molecule, the carbon atoms are arranged in a closed spherical shell. This allotrope became known as fullerene in tribute to the architect Fuller who designed intriguing constructions known as shaped polyhedrons. The fullerene, or $C_{60}$, is formed by twelve pentagonal rings combined with twenty hexagon rings of the carbon. It is well known in the theoretical and experimental setting \cite{xray,sheka},  that the X-ray diffraction pattern of  $C_{60}$ is well fitted when suggesting that the molecule shape is a truncated icosahedron of the point-group symmetry Ih.  The  carbon atoms are arranged in 20 six-membered and 12 five-membered rings. The C-C bonds separating two hexagons are double bonds ($h$) while the pentagon C-C bonds ($p$) are single \cite{sheka}.
As in the case of graphene, the conduction band of $C_{60}$ fullerene can be described by a tight-binding model \cite{Maria, Lammert e Crespi 1, Lammert e Crespi 2}. Therefore, the Fermi surface reduces to two $K$-points located in the Brillouin zone.

It is well known that the existence of pentagonal rings in $C_{60}$ fullerene can be interpreted as a disclination in the spherical lattice formed and described by a "cut and glue" Volterra process \cite{Volterra}.  
And these topological defects are associated with the symmetries during the generation of the molecule. The disclination becomes especially relevant in two-dimensional solids, being responsible for the value of the strength, and for the arising of curvature in the lattice. 
The geometric approach to its description is more powerful, because, in this theory, the disclinations are well described by a curvature of the lattice.
The curvature of the molecule induces an effective gauge field that is obtained via  the parallel transport of the  reference frames in a closed-circuit  around the defect~\cite{Lammert e Crespi 1,Lammert e Crespi 2,Claudio1,Birrel,CartroNetoGuineaPeres, VozmedianoKatsnelsonGuinea,Pachos}:
\begin{equation}
\oint \omega_{\mu}dx^{\mu}=-\frac{\pi}{6}\sigma^{3},
\label{quantumflux1}
\end{equation}
which enforces a mixture of the Fermi points ($\textbf{K}_{\pm}$) and generates a non-Abelian gauge field (or a K-spin flux) as demonstrated in \cite{Lammert e Crespi 1,Lammert e Crespi 2}, offsetting the discontinuity of the Bravais labels (A/B):
\begin{equation}
\oint A_{\mu}dx^{\mu}=\frac{\pi}{2}\tau^{2},
\label{quantumflux2}
\end{equation}
where $\tau^{2}$ is the second Pauli matrix mixing up the $\textbf{K}_{+}$ and $\textbf{K}_{-}$ components of the spinor on the $\textbf{K}_{\pm}$ space.
In other words, the mixture of Fermi points induces an effective field arising due to a fictitious magnetic monopole in the center of the fullerene by replacing the fields of 12 disclinations.

The first geometric continuous model using the Dirac equation to describe low-energy electrons in the $C_{60}$ fullerenes was proposed by Gonzales, Guinea and Vozmediano \cite{Maria}. Within this model, the fullerene molecule is described by a non-Abelian magnetic monopole introduced in the centre of a two-dimensional sphere. The field produced by this monopole represents the fields of twelve disclinations presenting in the $Ih$ (Icosahedron) symmetry. This field theory model was used by Kolesnikov and  Osipov~\cite{Osipov}, where pentagonal rings in fullerene  are simulated by two kinds of gauge fields. One of these gauge fields describes the elastic properties of disclinations, while the second one  is a non-Abelian gauge field describes the K-spin fluxes.

To study the properties of electrons taking place due to the presence of a
topological defect in graphene and $Ih$ fullerenes, it is natural to formulate the Dirac equation in a curved space induced by this defect.
However, here we will introduce Kaluza-Klein extension in the spherical metric in the next section.
\section{The extra dimension approach for $C_{60}$ fullerenes}\label{secIII}

Since the quantum flux in (\ref{quantumflux1}) can be obtained within the geometrical approach \cite{Claudio1}, a holonomy matrix can be calculated when the spinor suffers parallel transport around the apex of a graphitic cone.
The effect of the parallel transport of a spinor around the apex of the graphitic cone is analogous to the Aharonov–Bohm effect \cite{AharonovBohm}.
Recently, the geometrical description for a second quantum flux in graphene cones (\ref{quantumflux2}) was proposed in \cite{A Kaluza Klein description of geometric phases in graphene}, where the Fermi point degree of freedom, or K-spin flux, is described via a Kaluza-Klein theory.
The line element in the presence of a non-Abelian gauge field ($B_{\mu}$) is given in the form \cite{Love}
\begin{equation}
ds^{2}= g_{\mu \nu}(x)dx^{\mu}dx^{\nu} + (dy + \kappa B_{\mu}dx^{\mu})^{2},
\end{equation}
where $g_{\mu \nu}(x)$ is the metric tensor in (2 + 1) dimensions, and $y$ corresponds to the coordinate of the extra dimension ($\kappa$ is a constant called the Kaluza constant). 
Following the prescription of Refs. \cite{Love,Everton,A Kaluza Klein description of geometric phases in graphene}
for a two-dimensional spherical geometry, which describes the $C_{60}$ fullerene, we have:
\begin{equation}
ds^{2}= V_{f}^{2}dt^{2} - R^{2} d\theta^{2} - \alpha^{2}R^{2}\sin^{2}\theta d\phi^{2} - \bigg( dy + \frac{n_{\Omega}}{2} R \sin \theta d\phi \bigg )^{2}
\label{metrica}
\end{equation}
where $\alpha = 1 - \frac{n_{\Omega}}{6}$ is related to the angular sector removed from a spherical sheet in order to form the conical singularities, $V_{f}$ is the Fermi velocity of order $10^{6}$ m/s, and $R$ is the radius of fullerene, of the order of nanometers.
This effective geometry is considered by the fact pointed by Kolesnikov and  Osipov \cite{Osipov}: eigenfunctions of the low-energy levels oscillate relatively slowly with a distance. In this way, we can find the impact of the twelve pentagons considering the two-dimensional spherical metric near the defect.
Without the extra dimension, the element (\ref{metrica}) describes a sphere in (2 + 1) dimensions.

The bases of this space-time are known as tetrads (${e^{a}}_{\mu}(x)$), which are defined at each point in space-time by a local reference
frame $g_{\mu \nu}(x)=\eta_{ab}{e^{a}}_{\mu}{e^{b}}_{\nu}$ \cite{Birrel}. The tetrad and its inverse, ($e^{\mu}_{a}=\eta_{ab}g^{\mu \nu}e^{b}_{\nu}$), 
satisfy the orthogonal relationships: $e^{a}_{\mu}e^{b \mu}=\eta^{ab}$, $e^{a}_{\mu}e^{\mu}_{b}=\delta^{a}_{b}$, $e^{\mu}_{a}e^{a}_{\nu}=\delta^{\mu}_{\nu}$.
The Greek indices ($\mu, \nu$) for space-time frame coordinates, and the Latin indices ($a,b$) for local frame coordinates. For the local 
reference frame ($\theta^{a}=e^{a}_{\mu}(x)dx^{\mu}$), we choose the tetrads: 
\begin{eqnarray}
e^{a}_{\ \mu}(x) = \left(\begin{array}{cccc} V_{f} & 0 & 0 & 0 \\ 0 & R & 0 & 0 \\ 0 & 0 & \alpha R\sin\theta & 0 \\ 0 & 0 & \frac{n_{\Omega}}{2} R \sin \theta & 1  \end{array}\right);\ e^{\mu}_{\ a}(x) = \left(\begin{array}{cccc} \frac{1}{V_{f} }& 0 & 0 & 0 \\ 0 & \frac{1}{R} & 0 & 0 \\ 0 & 0 & \frac{1}{\alpha R\sin\theta} & 0 \\ 0 & 0 & -\frac{n_{\Omega}}{2\alpha} & 1
\end{array}\right).
\label{tetrads}
\end{eqnarray}

In a space with a curvature, the components of covariant derivative are given by
\begin{equation}
\nabla_{\mu}=\partial_{\mu}+\frac{i}{4}\omega_{\mu ab}(x)\Sigma^{ab}
\end{equation}
where $\Sigma^{ab}=\frac{i}{2}[\gamma^{a}, \gamma^{b}]$, and the $\gamma^{a}$ matrices are defined in the local reference frame and are identical to the Dirac matrices in the Minkowski spacetime in Weyl (or chiral) representation \cite{A Kaluza Klein description of geometric phases in graphene, GSC2007}.

The beautiful way to obtain the one-form connection (${{\omega}^{a}}_{b}$) is based on the first  Maurer-Cartan structure equation:
\begin{equation}
d\theta^{a}+{\omega^{a}}_{b} \wedge \theta^{b}=0 \mbox{.}
\label{Maurer-Cartan equation}
\end{equation}
where the non-zero components are given by:
\begin{eqnarray}
\omega^{\ 1}_{\phi \ 2}(x) = -\omega^{\ 2}_{\phi \ 1}(x) = -\alpha\cos\theta;\label{connection1}\\
\omega^{\ 1}_{\phi \ 3}(x) = -\omega^{\ 3}_{\phi \ 1}(x) = -\frac{n_{\Omega}}{2}\cos\theta.\label{connection2}
\end{eqnarray}
In this way, we can write the Dirac equation for a massless fermion in this background as
\begin{equation}
i\frac{\gamma^{0}}{V_{f}}\frac{\partial \psi}{\partial t} + i\frac{\gamma^{1}}{R} \bigg ( \frac{\partial}{\partial \theta} + \frac{\cot \theta}{2} \bigg ) \psi + i\frac{\gamma^{2}}{\alpha R \sin \theta}\frac{\partial \psi}{\partial \phi} - i \gamma^{2} \frac{n_{\Omega}}{2\alpha} \frac{\partial \psi}{\partial y} + i \gamma^{3} \frac{\partial \psi}{\partial y} - \frac{n_{\Omega}}{4 \alpha R}\cot \theta \gamma^{0}\gamma^{5} \psi =0.
\label{Dirac equation}
\end{equation}

Since the extra dimension is periodic, we can expand the Dirac spinor in the Fourier series with  respect to this dimension: 
\begin{equation}
\psi (t,\theta, \phi, y)=\sum_{l = - \infty}^{\infty} \exp{ \bigg ( 2\pi i l \frac{y}{L} \bigg )}\psi_{l}(t,\theta, \phi)
\end{equation}
where the scale of the extra dimension is denoted by $L$.
Now, the Dirac equation (\ref{Dirac equation}) takes the form
\begin{equation}
i\frac{\gamma^{0}}{V_{f}}\frac{\partial \psi_{l}}{\partial t} + i\frac{\gamma^{1}}{R} \bigg ( \frac{\partial}{\partial \theta} + \frac{\cot \theta}{2} \bigg ) \psi_{l} + i\frac{\gamma^{2}}{\alpha R \sin \theta}\frac{\partial \psi_{l}}{\partial \phi} 
- \frac{n_{\Omega}}{4 \alpha R}\cot \theta \gamma^{0}\gamma^{5} \psi_{l} +
 \frac{2\pi l}{L} \frac{n_{\Omega}}{2\alpha} \gamma^{2} \psi_{l} - \frac{2\pi l}{L} \gamma^{3} \psi_{l} =0.
\label{Dirac equation in Fourier modes} 
\end{equation}

Note that the last two terms in (\ref{Dirac equation in Fourier modes}) correspond to the massive modes created by the extra dimension of the KK theory.
Obviously, we can apply the Dirac phase factor method in (\ref{Dirac equation in Fourier modes}). 
We will do that in the next section.
\section{Quantum Holonomy}\label{secIV}

The possibility to realize a quantum gate based by a geometric description (Eq. \ref{metrica}) establishes a problem of obtaining the geometrical phase ($\Phi$) of the system in (\ref{Dirac equation in Fourier modes}). To do this, we must use the Dirac phase factor method, where we suppose that the Dirac spinor is written as
\begin{equation}
\psi_{l}(t, \theta, \phi)=e^{i\Phi}\psi_{l}^{0}(t, \theta, \phi) = \exp{ \bigg ( - \int \Gamma_{\mu}(x)dx^{\mu} \bigg )}\psi_{l}^{0}(t, \theta, \phi)
\end{equation}
which $\Gamma_{\mu}(x)=\frac{i}{4}\omega_{\mu a b}(x)\Sigma^{ab}$ being the spin connection. 
As we found, we can calculate the spinorial connection by defining a local reference frame at each point along the closed curse around the defect.
Thus the geometrical phase $\Phi$ defines the holonomy matrix: $U(\alpha, \theta)=e^{i\Phi}$, which stands here for a parallel transport of a spinor along a path around the defect are respectively
\begin{equation}
\Phi = \int \bigg ( \frac{\alpha \cos \theta}{2} \Sigma^{3} - \frac{n_{\Omega}}{4}\cos \theta \Sigma^{2} \bigg ) d\phi =
\pi \alpha \cos \theta \Sigma^{3} - 3\pi(1-\alpha)\cos \theta \Sigma^{2}
\label{fase}
\end{equation}
and
\begin{equation}
U(\alpha, \theta)=e^{i \pi \alpha \cos \theta \Sigma^{3} - 3i\pi(1-\alpha)\cos \theta \Sigma^{2}}.
\label{holonomia1}
\end{equation}

Here it is important to note that we can reduce the holonomy matrix (\ref{holonomia1}) to a single parameter $\alpha$ if we assume that $\theta$ is small, so, $\cos \theta \approx 1$. 
Remembering that parameter $\alpha =1-\frac{n_{\Omega}}{6}$ is related to the number of sectors removed ($n_{\Omega}$) in the creation of the defect in $C_{60}$ molecule. More about the synthesis of defects in graphene allotropes are found in \cite{Krishnan, Terrones, Sattler}.

Since equation (\ref{fase}) was obtained without adiabatic approximation, it's an Aharonov-Anandan phase. So that, in (\ref{holonomia1}) we have
\begin{equation}
U(\alpha)=e^{i \pi \alpha \Sigma^{3} - 3i\pi(1-\alpha)\Sigma^{2}}.
\label{holonomia2}
\end{equation}
This approximation  requires  to use $\rho =R \theta$ in the metric (\ref{metrica}). For the fullerene molecule, we consider $\rho \ll R$, where $\rho$ is the dimension of the local conical defect, and $R$ is the sphere radius. That way we have: $\sin \theta \approx \frac{\rho}{R}$ and $\cos \theta = \sqrt{1-\frac{\rho^{2}}{R^{2}}} \approx 1$.

Since the holonomy transformation (\ref{holonomia2}) involves the sum of two non-commuting matrices in the argument of the exponential function, we have that $e^{A+B}\ne e^{A}e^{B}$.
However, using the Hausdorff formula $e^{A+B}=e^{A}e^{B}e^{-\frac{1}{2}[A,B]}\dots$, and defining $\zeta=3(1-\alpha)$, we find
\begin{equation}
U(\alpha)\approx e^{i \pi \alpha \Sigma^{3}}e^{- i \pi \zeta \Sigma^{2}}e^{i\pi^{2}\alpha \zeta \Sigma^{1}}.
\label{holonomia3}
\end{equation}

By using the definition of the exponential of a matrix, that is, $e^{A}=\sum\limits_{i=0}^{\infty}\frac{A^{n}}{n!}$, the expression (\ref{holonomia3}) can be rewritten in the form:
\begin{equation}
U(\alpha)\approx q_{0}I+q_{1}i\Sigma^{1}-q_{2}i\Sigma^{2}+q_{3}i\Sigma^{3},
\end{equation}
where the parameters $q_{i}$ are defined as
\begin{eqnarray}
\begin{array}{ll}
q_{0}=\cos(\pi \alpha)\cos(\pi \zeta)\cos(\pi^{2}\alpha \zeta)+\sin(\pi \alpha)\sin(\pi \zeta)\sin(\pi^{2}\alpha \zeta), \\
q_{1}=\cos(\pi \alpha)\cos(\pi \zeta)\sin(\pi^{2}\alpha \zeta)-\sin(\pi \alpha)\sin(\pi \zeta)\cos(\pi^{2}\alpha \zeta), \\ 
q_{2}=\cos(\pi \alpha)\sin(\pi \zeta)\cos(\pi^{2}\alpha \zeta)+\sin(\pi \alpha)\cos(\pi \zeta)\sin(\pi^{2}\alpha \zeta), \\
q_{3}=\sin(\pi \alpha)\cos(\pi \zeta)\cos(\pi^{2}\alpha \zeta)-\cos(\pi \alpha)\sin(\pi \zeta)\sin(\pi^{2}\alpha \zeta). \\
\end{array} 
\label{parameters}
\end{eqnarray}
In this way the holonomy matrix can be described by an unitary matrix:
\begin{equation}
U(\alpha)\approx
\left( \begin{array}{cccc}
q_{0}+iq_{3} & -\left(q_{2}+iq_{1}\right)^{*} & 0 & 0 \\
q_{2}+iq_{1} & \left(q_{0}+iq_{3}\right)^{*} & 0 & 0 \\
0 & 0 & q_{0}+iq_{3} & -\left(q_{2}+iq_{1}\right)^{*} \\
0 & 0 & q_{2}+iq_{1} & \left(q_{0}+iq_{3}\right)^{*} \\
\end{array} \right) \mbox{.}
\label{quantum holonomy}
\end{equation}
It is also possible to show that $\sum q_{n}^{2}=1$. 
Also, the result we obtained in eq.  (\ref{quantum holonomy}) is very important because, as is well known,  quantum gates can be represented by unitary transformations applied at defined moments of time throughout the execution of the algorithm. So, any unitary transformation can be written as a combination of the elementary quantum logical keys \cite{Oliveira, Nielsen Chuang}. 

It is important to emphasize that continuum models elucidate a long-range physics at the same time as they are restricted to a few dozens of electron-volts around the Fermi point. Concurrent to this, the electronic density of states decrease in the neighborhood of the defect \cite{Lammert e Crespi 1, Lammert e Crespi 2}. 
Besides that, if we consider the $C_{60}$ molecule at low temperatures, we can reduce the thermal noise and the density of electrons on the conduction band.
In this way, the quantum holonomy could mixed up the $K_{\pm}$ Fermi points, and the electron location on sublattices $A$ and $B$. For this, we assumed that the sublattices were mixed when the control parameter changes into the range $0<\alpha <1$, and the gauge transformation on K-spin part is a rotation the components of a possible two-qubits spinor in the next step of the quantum computation issue.

\section{Conclusion}\label{secV}

Let us discuss our results. We have proposed a discussion about a quantum holonomy matrix based on the introduction of Kaluza-Klein extra dimension as an alternative manner to explain the mixture of Fermi points in the $C_{60}$ fullerene molecule with icosahedral ($Ih$) symmetry.
We have shown that it is possible to describe the mix of the Fermi points ($K_{\pm}$) within the $(4+1)$-dimensional geometrical approach. 
This result is important since it shows that the implementation of quantum gates by an extra dimension approach is an acceptable approximation, at least for the low-energy states, where the eigenfunctions of the quasiparticles inside the molecule do not oscillate too rapidly with distance.
Besides that, the adding of an extra dimension promote more freedom degrees on a possible two-qubits logical basis. That can be an EPR pair of entangled electrons, a pair electron-hole, or another one. 

It is hard to achieve the implementation of the quantum computation associated with topological defects in solids, even more with the use of extra dimensions. However, we wish to open a discussion about a new possibility to implement it.
\section{ACKNOWLEDGMENTS}

$\quad$We thank CAPES , CNPQ and FAPESQ-PB for financial support.



\end{document}